# Hairpin RF Resonators for MR Imaging Transceiver Arrays with High Inter-channel Isolation and $B_1$ Efficiency at Ultrahigh Field 7T

Komlan Payne[1], Leslie L. Ying[1,2] and Xiaoliang Zhang[1,2,3*]

[1]Department of Biomedical Engineering, State University of New York at Buffalo, Buffalo, NY 14260 USA
[2]Departments of Electrical Engineering, State University of New York at Buffalo, Buffalo, NY 14260 USA
[3]Univerity of California Berkeley and UCSF Joint Bioengineering Program, San Francisco, CA 94158 USA
**Correspondence author, xzhang89@buffalo.edu**

*Abstract*— **Electromagnetic decoupling among a close-fitting or high-density transceiver RF array elements is required to maintain the integrity of the magnetic flux density from individual channel for enhanced performance in detection sensitivity and parallel imaging. High-impedance RF coils have demonstrated to be a prominent design method to circumvent these coupling issues. Yet, inherent characteristics of these coils have ramification on the B1 field efficiency and SNR. In this work, we propose a hairpin high impedance RF resonator design for highly decoupled multichannel transceiver arrays at ultrahigh magnetic fields. Due to the high impedance property of the hairpin resonators, the proposed transceiver array can provide high decoupling performance without using any dedicated decoupling circuit among the resonant elements. Because of elimination of lumped inductors in the resonator circuit, higher B1 field efficiency in imaging subjects can be expected. In order to validate the feasibility of the proposed hairpin RF coils, systematical studies on decoupling performance, field distribution, and SNR are performed, and the results are compared with those obtained from existing high-impedance RF coil, e.g., "self-decoupled RF coil". To further investigate its performance, an 8-channel head coil array using the proposed hairpin resonators loaded with a cylindrical phantom is designed, demonstrating a 19 % increase of the $B_1^+$ field intensity compared to the self-decoupled coils at 7T. Furthermore, the characteristics of the hairpin RF coils are evaluated using a more realistic human head voxel model numerically. The proposed hairpin RF coil provides excellent decoupling performance and superior RF magnetic field efficiency compared to the "self-decoupled" high impedance coils. Bench test of a pair of fabricated hairpin coils prove to be in good accordance with numerical results.**

## I. INTRODUCTION

RADIO frequency (RF) loop coil or antenna mainly used for direction finding (DF) system, has also been used in biomedical imaging applications [1]. These efficient antennas used for magnetic resonance imaging (MRI) [2] are considered electrically large with their perimeter comparable to the Larmor frequency wavelength (usually about a tenth of the wavelength or more) [3]. Loop antennas can operate either as transmitter/receiver to reconstruct an image using methods namely root-sum-of-squares magnitude combining, $B_1$-weighted combining, and parallel imaging such as sensitivity encoding (SENSE) [4-8]. The quest for ultimate intrinsic signal-to-noise ratio (uiSNR) and higher spatial resolution has enabled the study of MRI at ultra-high field (UHF) where the static magnetic field $B_0 \geq 7T$. However, at UHF the effective wavelength of the Larmor frequencies is reduced and comparable to the anatomic dimension causing a significant distortion in the field distribution. The undesired effect of the electromagnetic wave propagating in the body tissues renders the field inhomogeneous [9-14]. Attention has been placed toward the architecture of the antennas within the MRI system to address this issue. While the birdcage coil [15, 16], the transverse electromagnetic (TEM) resonator [11, 17-20] and microstrip resonator [11, 21-24] are sources of homogeneous RF excitation field $B_1$ across their entire covered volume at low frequency, they have failed to provide such uniformity at UHF. RF array coils composed of several distributed surface coils owing to the near field pattern control can be used at UHF. By dynamically adjusting the excitation phase and amplitude of individual transmit elements through parallel excitation (pTx), a relatively uniform B1 field can be obtained in the region of interest (ROI) in humans. This method namely "B₁ shimming" provides a more localized high SNR at the ROI [13, 25, 26]. However, uiSNR and acceleration performance are contingent on a very low crosstalk power between individual elements of RF array coils. For high performance in detection sensitivity and imaging acceleration, a very close-fitting array is required. Such configuration leads to strong coupling among resonant elements of the array making the decoupling mechanism a very challenging task.

Mutual coupling in high-density arrays of antenna is intrinsically inevitable. For such, this interaction alters the isolated current distribution and impedance of each individual element [10, 27-33]. The magnetic flux linkage between RF coils through direct space coupling or scattering from nearby objects leads to correlated noise which degrades the SNR and the parallel imaging performance. Few methods have been explored in the past to circumvent strong coupling for RF coils array. One of the straightforward techniques is based on the geometry of overlapping decoupling for adjacent elements [34]. This method is usually combined with a low-impedance preamplifier to assure sufficient decoupling between non-adjacent elements; however, this method cannot be readily integrated into transceiver arrays [12, 35-39]. Furthermore, due to the overlapped area, unoptimized imaging performance is expected for accelerated parallel imaging techniques. Other techniques adopted include the use of transformers, capacitive/inductive decoupling lumped networks, and passive distributed resonators between inter-elements, i.e. magnetic wall decoupling circuits [27, 28, 31, 33, 40-46]. In addition, with the evolution of artificial periodic material commonly named as "frequency selective surface" [47], magnetic wall distributed filter and metamaterial superstrate are developed to suppress surface wave propagation between antennas in an array [48, 49]. Although these methods provide excellent decoupling performance, the overall design array becomes cumbersome when the number of coils increases which leads to a much more complicated synthesis.

High impedance coils (HIC) [34, 50-53], a decoupling design mechanism, do not require any additional circuit between resonant elements to provide excellent decoupling between adjacent coils. The physical insight behind this method is to induce electric coupling within a conventional balanced loop to cancel the inherent magnetic coupling in order to reduce the total flux density between neighboring elements. Due to the high impedance of the resonator circuit, an open-path current pattern can be created which is characteristic of an electric dipole field distribution with the ability to support both magnetic and electric modes. "Self-decoupled" RF coils [50], one of the prominent HIC have demonstrated good decoupling performance and image quality compared to overlapped conventional coils. However, this decoupling technique comes at the expense of the SNR and B1 field efficiency. In fact, the use of lumped inductor with a relatively large value of inductance is needed in the high impedance self-decoupled RF coils for maintaining the desired resonance frequency and sufficient decoupling. Due to their inherently low-quality factor, a significant amount of power is expected to be dissipated within these inductors (winding and core losses). Additionally, a considerable amount of B1 fields is stored in the lumped inductors and not readily available for MR imaging, leading to a degraded B1 efficiency in the imaging subject. Therefore, the design of the self-decoupled

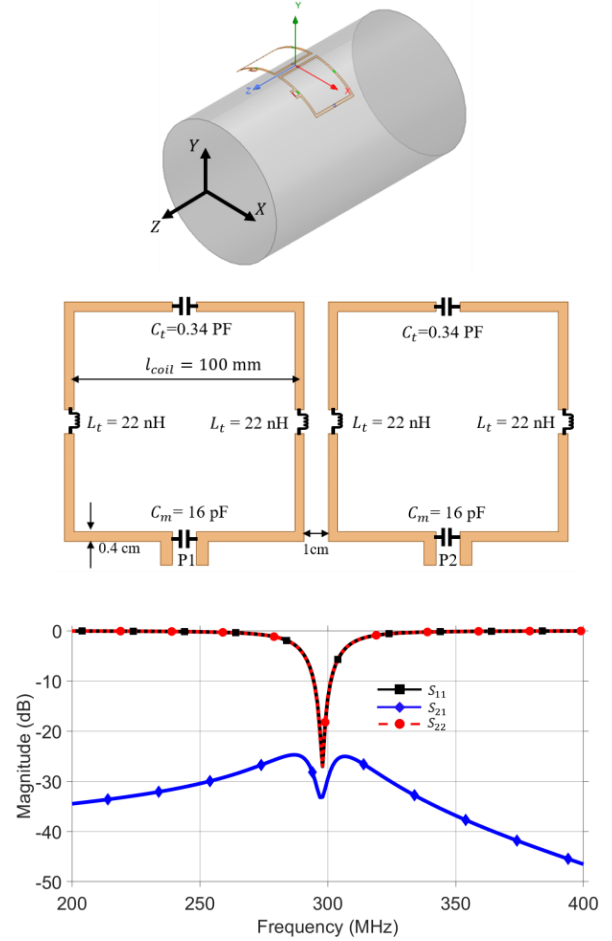

Fig. 1. (a) Topology of self-decoupled RF coils along with detailed geometry dimension and electrical parameters. Pair of loops on top of a cylindrical phantom (conductivity $\sigma = 0.6$ S/m and permittivity $\varepsilon_r = 50$). (b) Simulated scattering parameters of the self-decoupled RF coils. At the Larmor frequency, the reflection coefficient is about 28 dB (excellent input impedance matching), and the transmission coefficient is 33 dB (0.05 % power crosstalk).

coils has the limitation in reaching their uiSNR.

In this paper, we propose a hairpin RF coil design technique that provides excellent decoupling performance without compromising B1 efficiency. The design principle of operation is illustrated with its equivalent circuit analysis. Full-wave electromagnetic analysis using High-Frequency Structure Simulator (HFSS) from the ANSYS Corporation is performed to evaluate the performance of the proposed hairpin RF coil and compared to the self-decoupled coils at 7T. The results include decoupling performance, current and electric field distribution, RF field B1 maps, local specific absorption rate (SAR), and SNR field intensity. Bench test result of a pair of fabricated hairpin coils is also provided. In Addition, the attributes of the proposed hairpin RF coils are evaluated using a realistic build-in human head voxel model from ANSYS HFSS.

## II. DESCRIPTION OF METHODOLOGY

In this section, a pair of self-decoupled RF coils [50], is designed to operate at 298 MHz (Larmor frequency of 7T for proton [1]H). Another pair of self-decoupled RF coil with a larger size without lumped inductors is also designed to provide the same decoupling performance. Then their performances are compared with the proposed hairpin RF coils designed at the same operating frequency. For a fairly assessment of all the different type of RF coils, a 4 mm copper width (conductivity $\sigma = 5.87 \times 10^7$ S/m) is used for the conductor. All the designs are matched to 50 Ohms, the characteristic impedance of most MRI scanners equipment. The lumped capacitors and inductors are models as lossy series resonators with realistic quality factors. The Q-factor of the capacitors is set to 1000-1500 while the inductors (with large inductance value) Q-factor is set to 250, value taking from their datasheet.

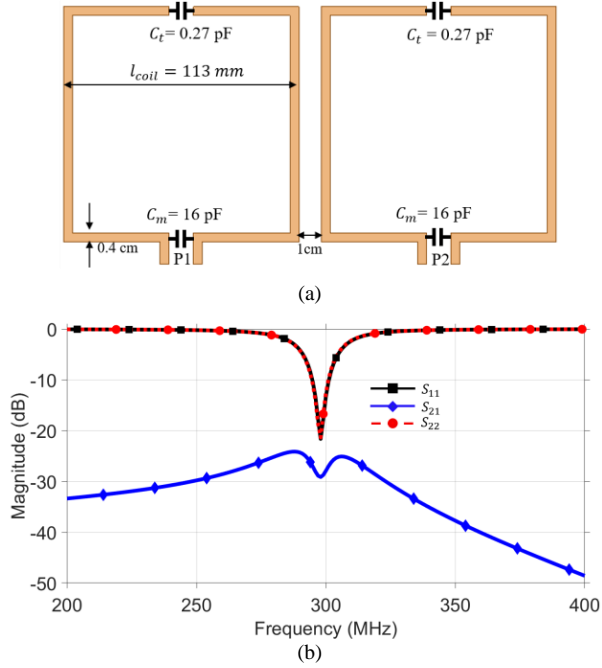

(a)

(b)

Fig. 2. Self-decoupled RF coils with larger size along with detailed geometry dimension and electrical parameters. Simulated scattering parameters of the RF coils showing excellent input impedance matching and good isolation between the pair of the RF coils.

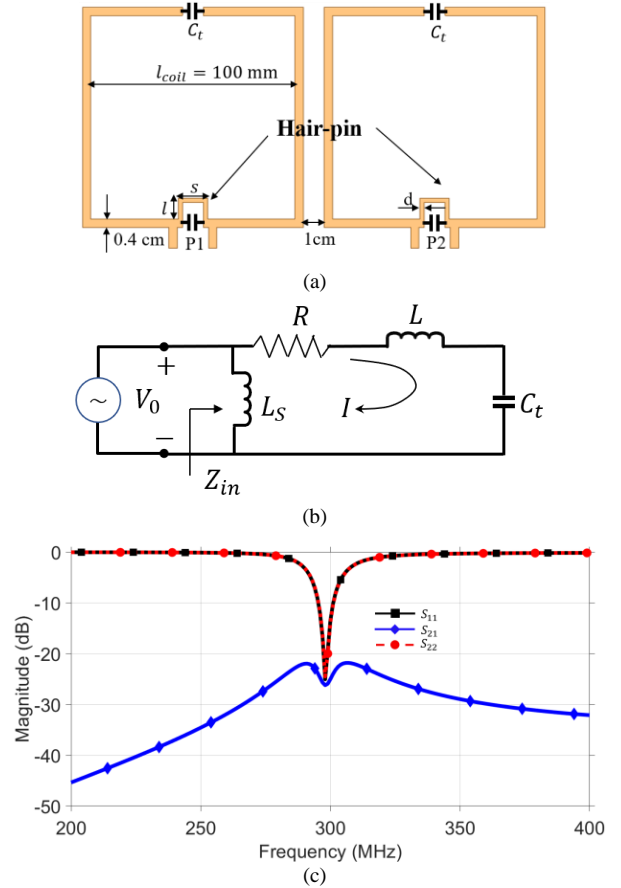

(a)

(b)

(c)

Fig. 3. Hairpin RF coils along with detailed geometry dimension and electrical parameters ($C_t = 0.4$ pF, $C_m = 5$ pF, $l = 1$ cm, $s = 1$ cm, d = 0.2 cm. Equivalent circuit model of individual coil. Simulated scattering parameters of the RF coils showing excellent matching and isolation performance between the pair of the RF coils. The transmission coefficient is about 26 dB (0.25 % power crosstalk).

### A. Self-decoupled RF coils for 7T

A pair of self-decoupled RF coils is modeled in ANSYS HFSS following the description provided in [50]. Each coil, a square loop with $10 \times 10$ cm$^2$ dimension is segmented at the center of each side of the loop with a lumped element. The two loops are separated from each other with 1 cm gap and wrapped on a 16.5 cm radius virtual cylindrical surface. As shown in Fig. 1(a), the coils are placed 4 cm on top of a cylindrical phantom (conductivity $\sigma = 0.6$ S/m and permittivity $\varepsilon_r = 50$) to mimic the human brain tissue. The input port of the first and second coil is denoted by P1 and P2, respectively. The radius and length of the cylindrical phantom are set to 12.5 cm and 30 cm, respectively. The simulated scattering parameters versus frequency in Fig. 1(b), shows excellent input impedance matching and strong decoupling between the adjacent coils. However as described in the previous section, a large value of inductance ($L_t = 22$ nH) is required for intrinsic decoupling.

In order to avoid the lumped inductors within the design, one can increase the size of the loop to compensate for the effective inductance. The self-decoupled RF coil with a larger size ($11.3 \times 11.3$ cm$^2$) without lumped inductors is designed as shown in Fig. 2(a). The simulated coupling between the adjacent coils observed in Fig. 2(b) is about -29 dB. Although, excellent decoupling performance is achieved with the larger size coils design, this is not a good candidate choice for the design improvement due to the filling factor since compact array of small RF coils generally provides higher SNR intensity [54]. An alternate HIC RF coil without the need of larger inductance value or larger coil size is tailored for high SNR intensity.

### B. Hairpin RF coils for 7T

We proposed a pair of hairpin coils without the need for the lumped inductance while keeping the coil size the same (see Fig. 3a). In order to account for the effective inductance in each loop to preserve intrinsic decoupling, a small conductor wire is connected across the driven terminal of each coil. The wire can take any shape form (in this case, a U-shape is adopted) and can be viewed as inductance across the feeding of the loop used to adjust the surface impedance of the coil. The hairpin length in the order of $\lambda/100$ (where $\lambda$ is the wavelength of the Larmor frequency) is a non-radiating line and has inductive reactance. To elucidate the working principle of the hairpin incorporated in the HIC coils, the equivalent circuit model (ECM) of an individual coil is shown in Fig. 3(b). The radiation resistance and the distributed inductance of the coil is denoted by R and L, respectively; the tuning capacitance is denoted as $C_t$; the hairpin is modeled as inductive shunt ($L_s$) across the coil feed point. The reactance value of the hairpin can be approximated as a shorted end transmission line [55]:

$$X_{L_s} = Z_0 \tan(2\pi l/\lambda) \qquad (1)$$

where $Z_0$, the characteristic of the two-parallel line of the hairpin is expressed as follow:

$$Z_0 = 276 \times \log_{10}(2s/d) \qquad (2)$$

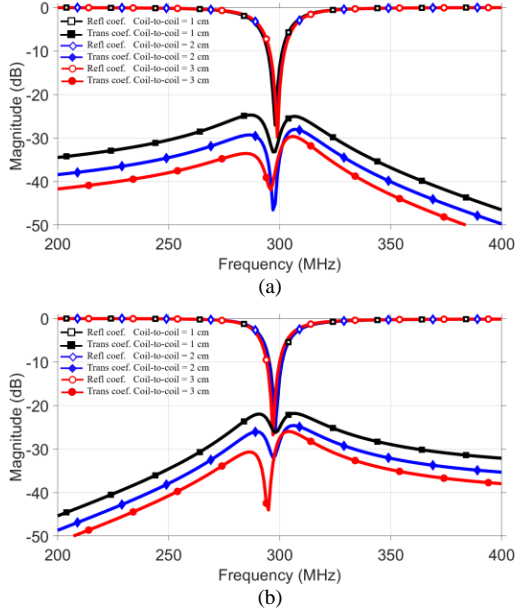

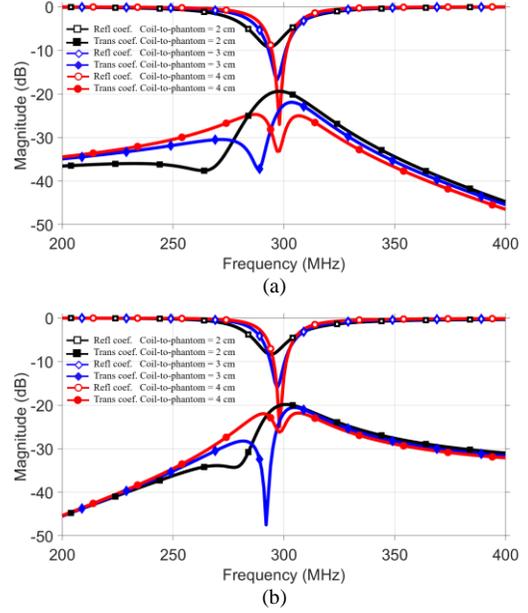

Fig. 4. Simulated scattering parameters of the RF coils subjected to different coil-to-coil separation (from 1 cm to 3 cm). (a) Self-decoupled coils; (b) Hairpin RF coils. Both designs demonstrate firm robustness to coil-to-coil separation.

Fig. 5. Simulated scattering parameters of the RF coils subjected to different coils-to-phantom separation (from 2 cm to 4 cm). (a) Self-decoupled coils; (b) Improved self-decoupled coils. High isolation between coils is maintained. However, both designs are sensitive to heavy loading.

The input impedance of the individual coil is therefore altered by the hairpin line and is expressed as:

$$\frac{1}{Z_{in}} = \frac{1}{jX_{L_s}} + \frac{1}{Z'_{in}} \qquad (3)$$

$$Z'_{in} = R + j\omega L + \frac{1}{j\omega C_t}. \qquad (4)$$

The simulated scattering parameters versus frequency (see Fig. 3(c)) of the hairpin RF coils using the parameters depicted in the caption of Fig. 3, show excellent matching performance ($S_{11}$ = -25 dB) for maximum transfer of power and strong decoupling performance ($S_{21}$ = -26 dB) for minimum power crosstalk between the pair of RF coils. The sensitivity of the frequency response obtained for both HIC design (with the same size) is evaluated against various distances between the pair of coils and the loading effect. The coil-to-coil distance is adjusted from 1 cm to 3 cm, while the coils-to-phantom distance varies from 2 to 4 cm. Both designs are initially optimized and matched at 298 MHz for 1 cm coil-coil separation and 4 cm coils-to-phantom separation. The frequency response seen in Fig. 4 shows that the performance of both designs is very robust to coils separation. Excellent input impedance matching, and high isolation are maintained between the pair of coils when the coil-to-coil separation changes from 1cm to 3 cm. The effect of various coils-to-phantom on the RF coils is illustrated in Fig. 5. The simulated results obtained show that high isolation between the pair of RF coils is maintained, however the design performance is sensitive to heavy loading. As can be seen from both designs, the resonant frequency has shifted from 298 MHz to 293 MHz and there is a degradation of the matching performance as the load is moving closer to the RF coil.

### C. Bench test measurement of the hairpin RF coils

A pair of hairpin design board is manufactured with the same dimension used in the numerical simulations. The coils are made with 1 oz copper on a thin Teflon substrate. Ceramic fixed and trimmer capacitors (from Johanson Manufacturing) are integrated in the coils design for matching and tuning as show in Fig. 6(a). The coils are designed to operate at 7T Larmor frequency and matched to 50 ohms. The pair of hairpin coil is placed 1 cm on top of a tank phantom with roughly the same electrical parameter of the phantom used in the numerical simulation. The scattering parameters of the coils is obtained using a vector network analyzer (ZVL, Rohde & Schwarz, Munich, Germany) within a broadband frequency range. The measured S-parameters obtained for very closed spaced hairpin coil (coil-coil distance set to 1 cm) is shown in Fig. 6(b). Excellent impedance matching (< -25 dB) and good isolation (about -23 dB) are achieved between both ports.

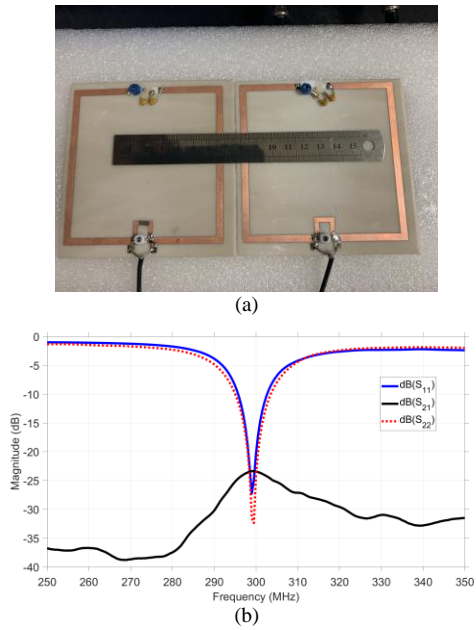

(a)

(b)

Fig. 6. (a) Photograph of a pair of the fabricated hairpin RF coils separated by 1 cm distance from each other. (b) Measured scattering parameters of the RF coils. Excellent matching (less than -25 dB) and good isolation (about -23 dB) are achieved between ports.

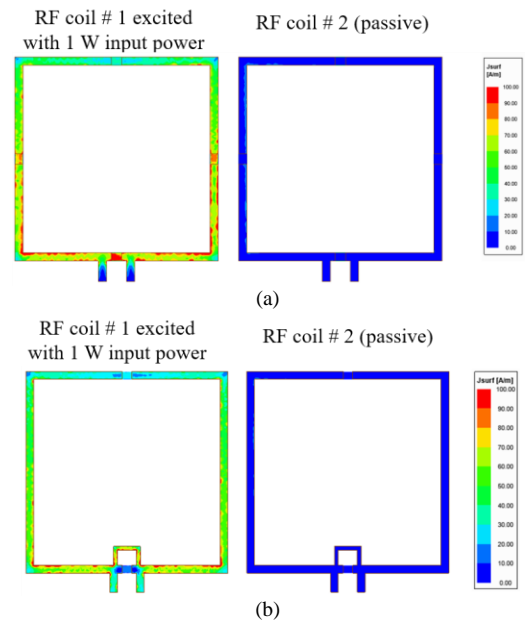

Fig. 7. Simulated surface current distribution obtained at 298 MHz. (a) Self-decoupled coils; (b) Hairpin RF coils. Both designs present practically the same current distribution along the conductors of the RF coils.

### D. Assessment of both HIC RF coils at 7T

The performance of the pair of self-decoupled coils with lumped inductors is identical to the proposed hairpin design in term of isolation at 7T. The evaluation of both designs in term of current and electric field distribution, RF transmit/receive ($B_1^+/B_1^-$) field strength maps, local SAR, and SNR field intensity is described below.

In conventional loop coil design, a uniform current distribution is obtained with equal values of lumped capacitors equally placed along the perimeter of the loop. Unlike conventional loop coil, different values of capacitance are utilized in self-decoupled coil design to deliberately produce a non-uniform current distribution. The simulated current density distribution of the self-decoupled RF coils at the Larmor frequency is illustrated in Fig. 7 where only one coil among the pair is excited with 1 W input power. As can be seen in both designs, the current is non-uniform along the conductor length of the loops (weak current is observed at the leg of the excited loop opposite to the feeding). Also, the residual current in the passive coils due to coupling between adjacent coils is very negligible, characteristic of strong decoupling performance for both designs. In a receive channel array, the local electric field from each channel is associated with the correlated noise between each pair of coils used in the calculation of SNR. Also, the induced electric field through

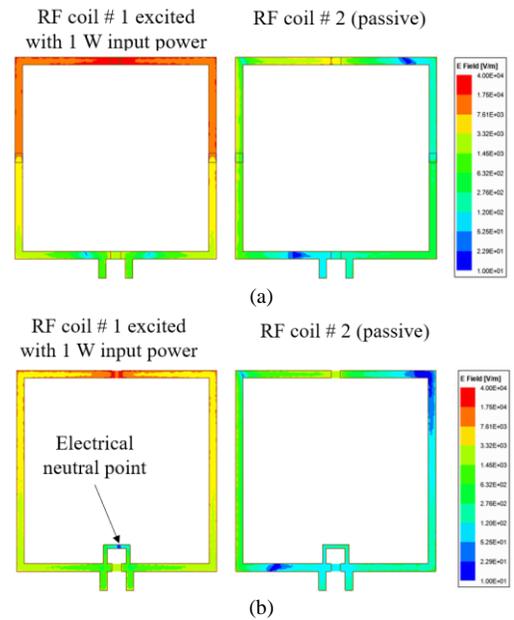

(a)

Electrical neutral point

(b)

Fig. 8. Simulated electric field distribution along the RF coils at 298 MHz. (a) Self-decoupled coils; (b) Hairpin RF coils. Overall, higher electric field amplitude is observed within the self-decoupled coils compared to the hairpin design.

the phantom from each coil in a transmit channel array is used for the calculation of the SAR. Hence, it is critical to explore the behavior of the electric field in the MRI system. The electric field distribution of the self-decoupled RF coils is illustrated in Fig. 8. A strong localized field enhancement is observed along the leg of the loop opposite to the feeding for both designs. Nevertheless, it can be seen the overall electric field strength in the self-decoupled coils is higher than the one induced in the proposed hairpin design.

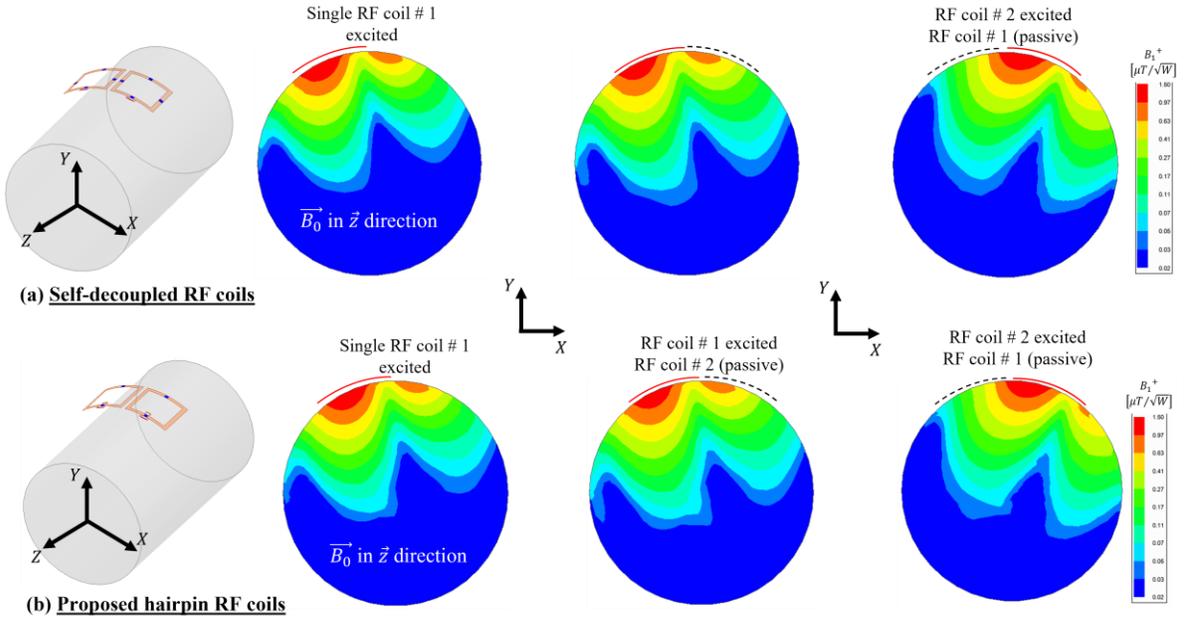

Fig. 9. Simulated transmit $B_1^+$ field distribution normalized to the accepted input power in the XY (transversal) plane through the center of the RF coils. Similar field distribution with insignificant coupling effect is obtained for both designs.

TABLE I. SIMULATED SCATTERING PARAMETERS OF THE 8 X 1 SELF-DECOUPLED ARRAYS RF COILS

| 1 | -39 | -23 | -30 | -44 | -36 | -43 | -30 | -23 |
| 2 | -23 | -19 | -23 | -30 | -44 | -37 | -43 | -30 |
| 3 | -30 | -23 | -41 | -23 | -30 | -43 | -36 | -44 |
| 4 | -44 | -30 | -23 | -41 | -23 | -30 | -43 | -36 |
| 5 | -36 | -44 | -30 | -23 | -43 | -23 | -31 | -44 |
| 6 | -43 | -36 | -43 | -30 | -23 | -38 | -21 | -31 |
| 7 | -30 | -43 | -36 | -43 | -31 | -21 | -20 | -23 |
| 8 | -23 | -30 | -44 | -36 | -44 | -31 | -23 | -45 |
| | 1 | 2 | 3 | 4 | 5 | 6 | 7 | 8 |

TABLE II. SIMULATED SCATTERING PARAMETERS OF THE 8 X 1 HAIRPIN ARRAYS RF COILS

| 1 | -21 | -25 | -39 | -38 | -40 | -39 | -39 | -20 |
| 2 | -25 | -19 | -21 | -37 | -39 | -41 | -38 | -37 |
| 3 | -39 | -21 | -18 | -20 | -39 | -39 | -40 | -39 |
| 4 | -38 | -37 | -20 | -19 | -22 | -36 | -38 | -41 |
| 5 | -40 | -39 | -39 | -22 | -19 | -24 | -37 | -38 |
| 6 | -39 | -41 | -39 | -36 | -24 | -19 | -20 | -36 |
| 7 | -39 | -38 | -40 | -38 | -37 | -20 | -18 | -22 |
| 8 | -20 | -37 | -39 | -41 | -38 | -36 | -22 | -19 |
| | 1 | 2 | 3 | 4 | 5 | 6 | 7 | 8 |

Another important EM near-field that required close attention in the study of MR imaging is the oscillating magnetic field B$_1$. For RF array coils, an ideal B$_1$ field map where the field profile from the individual element is distinct from neighboring coils is very crucial for parallel imaging. A good decoupling performance between the elements of the array is the key feature for an artifact-free map. However, the decoupling method should not come at the expense of the B$_1$ field strength. For such, we evaluate the B$_1$ field strength for both HIC RF coils designs. For UHF, there is a clear distinction between the transmit RF field ($B_1^+$) and the induced field caused by the precession phantom magnetization called the received field ($B_1^-$). Fig. 9 shows the simulated transmit $B_1^+$ field distribution normalized to 1 W of the accepted input power for both designs in the transversal plane cutting through the center of the RF coils. The red lines and black dashed lines are shown to allocate the location of the active and passive loops, respectively relative to the cylindrical phantom. The field distribution of the pair of element arrays is compared to the field generated by their counterpart single coil. It can be seen that the field map of only a single coil has a selective profile and is very similar to the one obtained from the element in the array with a negligible coupling effect. Both HIC designs also present similar $B_1^+$ field distribution.

To further highlight the image quality of the proposed hairpin RF coils over the existing self-decoupled coils, we evaluate the SNR of an array of 8 channels (with 1 cm circular gap between adjacent elements) wrapped on a virtual cylindrical surface of 14.5 cm radius. The coils are placed 2 cm on top of the cylindrical phantom of 12.5 cm radius and 30 cm length. The arrangement of the array over the phantom is illustrated in Fig. 10. All the 8 elements are excited using 50 ohms lumped ports with 1 W input power and 45° linear phase progression. The scattering parameter matrix of the 8 channels for both designs is depicted in Tables I and II. The simulated results obtained show good matching performance ($S_{ii} \leq$ -18 dB for all the channels) and excellent decoupling performance between all neighboring elements ($S_{ij} \leq$ -20 dB). The simulated SNR maps of the 8 channels array on the cylindrical phantom is calculated using the built-in Fields Calculator module in ANSYS HFSS. For comparison between both designs, we evaluate the combined transmit $B_1^+$ field distribution normalized to 1 W of the accepted input power and also the relative value of SNR defined as $SNR_{90} = SNR/\sin \alpha$, where the flip angle $\alpha = 90°$. The simulated field maps on the phantom obtained from the 8 elements arrays is illustrated in Fig. 10. It can be seen that a significant field boost is obtained for the proposed hairpin RF coils at the peripheral as well as the center of the phantom. The local SAR averaged over 10 g is also computed. Similar local SAR map within the phantom is obtained for both HIC designs and their peak value is about 2.5 W/Kg. A one-dimensional profile of the combined transmit $B_1^+$ field distribution normalized to 1 W of the accepted input power for both HIC designs at depth points (along the y-axis at the center of coil #1) inside the phantom is illustrated in Fig. 11. The quantitative data shows a gain in an average of 19% in $B_1^+$ field intensity for the proposed design array coils over the self-decoupled coils in the ROI. As predicted, the lower RF transmit $B_1^+$ field efficiency from the self-decoupled design is mostly due to the power lost within the lossy inductors.

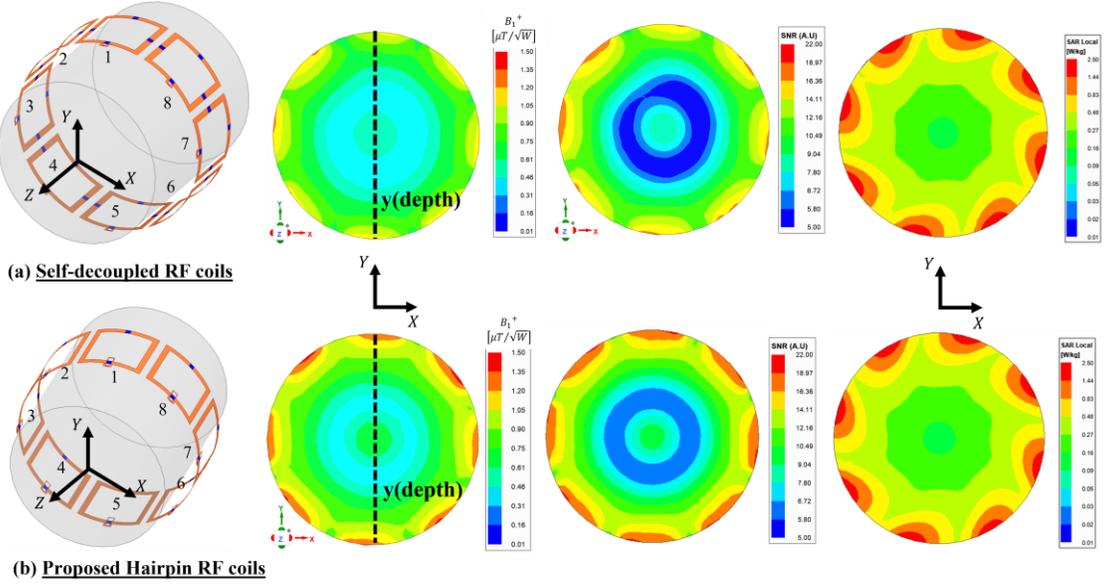

Fig. 10. Simulated of the combined transmit $B_1^+$ field distribution normalized to 1 W of the accepted input power, relative signal-to-noise ratio ($SNR_{90}$) map and local $SAR_{10g}$ on the phantom obtained from the 8-element arrays excited with 1 W input power and 45° linear phase progression.

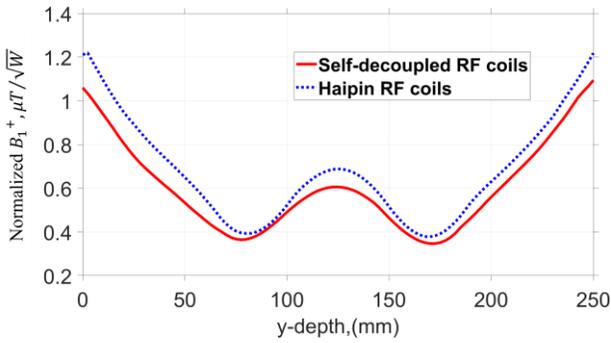

Fig. 11. One dimensional profile of the simulated $B_1^+$ per 1 W accepted power along the dash line (see inset of Fig. 10). The quantitative data shows a gain in an average of 19% in $B_1^+$ field intensity for the proposed hairpin RF coils over the self-decoupled coils.

## III. Simulation For Human Head Using 8 element Arrays Of The Proposed Hairpin RF Coils

For a more realistic analysis, we perform the simulations using an 8-element array of the hairpin RF coils described in Section II using a build-in human head voxel model as the phantom from ANSYS HFSS. The arrangement of the head voxel with the coils is illustrated in Fig. 12(a). The circular gap between adjacent coils is about 0.5 cm (very close fitting) and the distance between the head voxel model and the closest coil from the array is between 5-22 mm. The asymmetry of the human head voxel model and the close-fitting configuration of the design reflect on the design performance. Once Again, all the 8 elements are excited using 50 ohms lumped ports with 1 W input power and 45° linear phase progression. The scattering parameter matrix of the 8 channels is depicted in Fig. 12(b). The simulated results obtained show good matching performance ($S_{ii} \leq$ -18 dB for all the channels) and acceptable decoupling performance between adjacent and all neighboring elements ($S_{ij} \leq$ -14 dB). The simulated RF transmit/receive ($B_1^+/B_1^-$) field efficiency, SNR intensity map and local SAR are shown in Fig. 12(c). It can be seen that the ultimate intrinsic SNR are captured in various ROI of the human head being imaged.

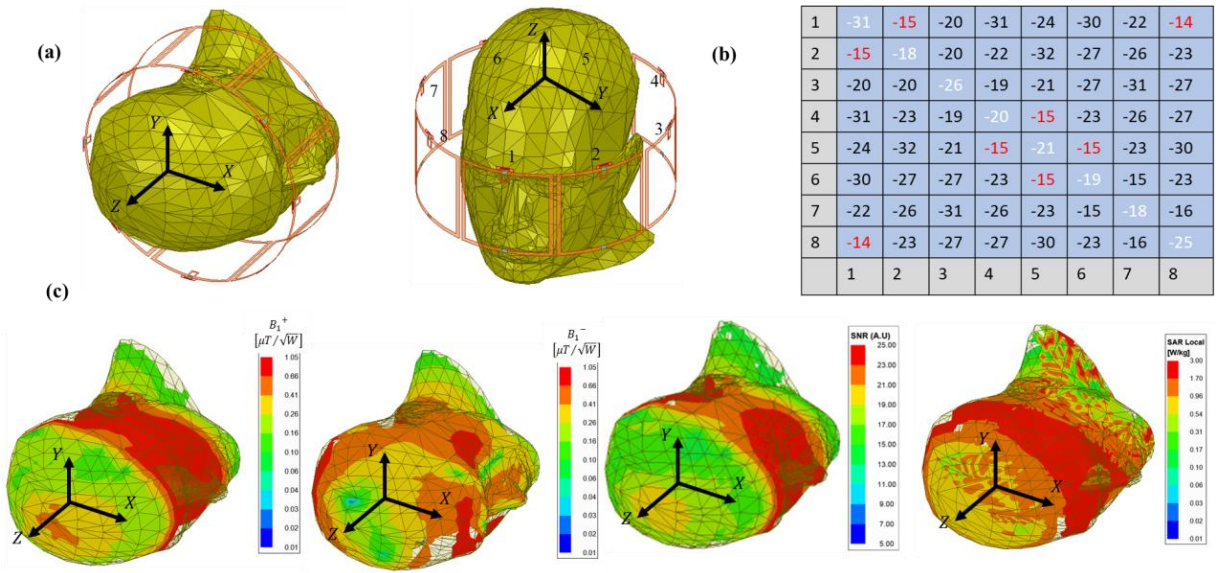

Fig. 12. Simulated transmit/receive ($B_1^+/B_1^-$) 3D field distribution, relative signal-to-noise ratio ($SNR_{90}$) map and local $SAR_{10g}$ on the head voxel model obtained from the 8-element arrays excited with 1 W input power and 45° linear phase progression.

## IV. Conclusion

We have proposed a high impedance RF coil design using the hairpin technique for multichannel transceiver coil arrays at the ultrahigh field of 7 Tesla. The proposed hairpin RF coils provide intrinsic decoupling performance without the need of the lossy lumped inductors in the resonator which compromises $B_1$ efficiency. As expected, this method demonstrated a substantial gain in the transmit efficiency $B_1^+$ essential for parallel imaging performance. The study shows that the proposed hairpin resonator is able to circumvent the significant $B_1$ field and SNR intensity losses from the self-decoupled coils. Although the technique is examined for 7T, it can be scaled to lower fields (static magnetic field $B_0 < 7T$). The proposed technique is expected to be a more general and robust solution to addressing the challenges of electromagnetic coupling and detection sensitivity encountered in high frequency RF transceiver arrays.

### Acknowledgment

This work is supported in part by the NIH under a BRP grant U01 EB023829 and SUNY Empire Innovation Professorship Award.